%% file: main.tex
\documentclass[authoryear,review,3p,times]{elsarticle}
\usepackage{amssymb}
\usepackage{amsmath}
\usepackage{bm} 
\usepackage{longtable}
\usepackage{mathtools}
\usepackage{multirow}
\usepackage{caption}
\usepackage{soul}
\usepackage{color}

\usepackage{array}

\journal{Journal of Computational Physics}

\begin{document}

\begin{frontmatter}



\title{The Method of Simultaneous Solutions Applied to Neutron Transport and Heat Conduction}


\author[MIT]{Dean Price\corref{cor1}}%% Author name
\ead{dnprice@mit.edu}
\author[uMich]{Brian Kiedrowski} %% Author name
\author[MIT]{Benoit Forget} %% Author name

\affiliation[MIT]{organization={Nuclear Science and Engineering, Massachusetts Institute of Technology},%Department and Organization
            addressline={60 Vassar St}, 
            city={Cambridge},
            postcode={02139}, 
            state={MA},
            country={US}}
\affiliation[uMich]{organization={Nuclear Engineering and Radiological Science, University of Michigan},%Department and Organization
            addressline={2355 Bonisteel Boulevard}, 
            city={Ann Arbor},
            postcode={48109}, 
            state={MI},
            country={US}}

%% Abstract
\begin{abstract}
This paper provides an initial description of the Method of Simultaneous Solutions, a Monte Carlo approach that simultaneously solves multiple Boltzmann-transport-like phenomena.
Here, it is used to simultaneously solve the neutron transport and heat conduction equations.
Analytically-derived weighting factors are tracked through a neutron transport-governed random walk to tally statistical estimators that can be used to calculate the temperature distribution.
In this initial presentation, the method is readily applicable to neutron-heat multiphysics problems where the heat source and neutron source are identically distributed spatially.
The primary theoretical benefit of MOSS lies in the reduction of computational cost that occurs from the removal of a dedicated routine to solve the heat conduction equation.
Practically, branching processes required to capture the disparate boundary conditions associated with these separate physical phenomena can lead to large computational times dedicated to a single physics.
In addition, this correlated sampling-based method can suffer from infinite variance associated with statistical estimators if the stochastic processes being tracked are too different.
The final drawback demonstrated in this paper is that the approximation of heat conduction as a Boltzmann transport-governed process leads to errors in calculated temperatures.
The paper explores these drawbacks on two demonstration problems, a problem consisting of slab geometry and a problem consisting of a hexagonal pin cell.
\end{abstract}


\begin{highlights}
\item Introduces the Method of Simultaneous Solutions (MOSS), a Monte Carlo technique that reuses neutron transport histories to estimate temperature fields with correlated sampling.
\item Formulates heat conduction as a purely scattering Boltzmann transport process with boundary-condition homogenization and albedo-based boundary treatment.
\item Demonstrates MOSS on slab and hexagonal pin-cell reactor problems, comparing neutron flux and temperature fields against reference solutions.
\end{highlights}

%% Keywords

\end{frontmatter}



\renewcommand{\hl}[1]{#1}

\section{Introduction}
\label{sec:intro}
Multiphysics analysis typically requires the simulation of several distinct physical processes that come together to form the total behavior of a system.
Although not realized in this work, its most complete form accounts for the interacting components of the physical processes for a fully coupled, self-consistent solution.
For example, in nuclear reactors, neutron transport and heat transfer are two such interacting physical processes.
This paper presents a Monte Carlo method that simultaneously solves the neutron transport and conductive heat transfer problems, using an approximate stochastic model of heat conduction derived from Boltzmann particle transport.
This simultaneous solution is accomplished with analytical path weights that let a single set of random histories track separate physical phenomena.
This method will be referred to as the Method of Simultaneous Solutions (MOSS) and can potentially be used to build tightly coupled multiphysics methods entirely within a Monte Carlo framework.

The benefits of using the Monte Carlo method for solving the Boltzmann transport equation are well-established and include the ability to represent complex geometries without discretization in the spatial, angular or spectral dimensions.
The most universal drawback comes from its low order of convergence and associated high computational cost.
However, other practical drawbacks arise when using these codes in multiphysics frameworks.
Deterministic approaches benefit from widely applicable numerical methods such as the finite element method that can be applied similarly to distinct physical phenomena.
In a general sense, this consistency in the methodologies makes multiphysics coupling more straightforward and even enables tight coupling where a single ``nonlinear solver simultaneously drives down the residual for each physics, taking into account the coupling between the physics in each nonlinear iteration'' \citep{novascone2013results}.
Monte Carlo methods, on the other hand, tend to be specialized and designed to treat a specific physical phenomenon. 
The other practical drawback observed when incorporating Monte Carlo Boltzmann equation solvers is that the constructive solid geometry representations of the problems can lead to consistency issues with typical mesh-based methods.
The present method, MOSS, seeks to address some of these issues.
It uses a consistent mathematical approach and geometric representation for Monte Carlo neutron transport and heat conduction.

MOSS can rely on the approximation of the diffusion equation used to model heat conduction with the Boltzmann transport equation.
The relationship between the Boltzmann transport equation and the diffusion equation is well-explored in the context of approximating transport with diffusion \citep{habetler1975uniform, papanicolaou1975asymptotic, dematte2024equilibrium}.
Specifically, \citet{larsen1974asymptotic} demonstrates that the leading order term in an asymptotic expansion of the transport solution is governed by the diffusion equation.
\citet{bensoussan1979boundary} \hl{later provided a fully rigorous derivation of this diffusion equation limit for linear transport.}
The control of this leading order behavior on the solution is ultimately dictated by the size of the particle mean free path.
Of course, the subject of interest in the current paper is reversing the direction of this approximation, representing diffusion with transport.
Fortunately, this direction of approximation allows for the introduction of a manufactured parameter, $\beta$, to increase the dominance of this leading order solution by lowering the effective mean free path of the ``heat particle''.
This increases the accuracy of the diffusion to transport approximation but, within the context of MOSS, can increase the variance in statistical estimators.

Independent of the transport equation, Monte Carlo methods for solving diffusion-based problems have been presented in the past \citep{muller1956some, haji1965application, bahadori2018mesh}.
However, MOSS relies specifically on the transport approximation for solving heat conduction to allow for consistency with the solution method for the neutron transport component of the problem.
\citet{Fraley1977MonteCarlo} initially presents the transport approximation to the heat conduction problem with additional work presented in \citet{fraley1980monte}.
Further exploration on the same method is presented later in \citet{Jun2007MonteCarloHeat}, \citet{song2007improved} and  \citet{cho2010monte}.
These studies tend to highlight the geometric flexibility and compatibility with existing Monte Carlo simulation tools as the main motivations for their work.
There is little complication in the treatment of the governing equations beyond the introduction of the $\beta$ parameter for scaling the mean free path.
Instead, as will be discussed in Section \ref{sec:bch}, the boundary conditions imposed on the transport equation in order to approximate those found in diffusion problems are not entirely straightforward.
In fact, as discussed in \citet{Jun2007MonteCarloHeat}, the need for boundary condition homogenization leads to an associated source-free diffusion problem that requires separate treatment.
In \citet{cho2010monte}, a different method for addressing this challenge is presented.

A central observation motivating MOSS is that in a fission-driven system, assuming local energy deposition, fission events are the common cause of both fields of interest.
They produce the neutrons that determine the flux and simultaneously produce heat that determines the temperature.
MOSS can exploit this common source because each simulated neutron history also seeds a co-located heat particle history governed by a similar transport process.
With this observation, the best application scenarios for MOSS involve those driven by fission reactions.
For now, MOSS is presented with a fixed-source form of the transport equation.

With both problems governed by a transport process, a technique called correlated sampling enables neutron histories already tracked for the neutron flux to be adjusted with weight factors to estimate the temperature field.
From \citet{rief1984generalized}, these weight factors are calculated with ``simple analytical expressions which can be computed with little additional effort''.
Unfortunately, a technique called branching is required to treat the boundary conditions associated with the heat conduction problem, this leads to some additional computational effort dedicated exclusively to solving for the temperature distribution.

The objective of the current work is to provide an initial presentation of MOSS with two demonstration examples.
%The novelty in this work lies in the recognition that fission events are the same source for neutrons and heat, MOSS reuses each sampled event to drive both the neutron walk and a $\beta$-scaled transport walk for heat transfer, enabling geometry-consistent tallies of temperature during the flux solve.
The novelty in this work lies in the recognition that the particle histories arising from a  Boltzmann transport process can be used to track a $\beta$-scaled transport process for heat conduction.
In its current form, MOSS has a few clear drawbacks compared to alternative methods for solving the heat transfer problem.
First, it introduces biases into the calculated temperature distributions due to the approximation of heat conduction using a Boltzmann transport process.
Next, the branching required to properly capture heat transfer boundary conditions does introduce some computational time that needs to be dedicated strictly to the heat transfer problem.
Finally, it requires separate treatment of a source-free diffusion problem to homogenize the boundary conditions.
Despite these limitations, the leveraging of existing neutron transport histories and geometric consistency  offers unique advantages that warrant further study.

The remainder of the paper is organized as follows, Section \ref{sec:method} provides a mathematical discussion of MOSS.
Section \ref{sec:transpssort} and Section \ref{sec:transeqapprox}, present the application of the Boltzmann transport equation to model neutron flux and temperature distributions.
Section \ref{sec:sscorsm} discusses the correlated sampling method and its role in MOSS.
Section \ref{sec:branching} and Section \ref{sec:clt_vio} discuss the drawbacks of MOSS in more detail.
Two demonstration problems are presented in Section \ref{sec:resultss} that consist of a slab geometry and pin cell geometry.
These demonstration problems are used to further discuss some aspects of MOSS including the computational cost associated with particle branching and errors from approximating heat conduction with the Boltzmann transport equation.

\section{Methodology}
\label{sec:method}

\subsection{Transport Equation for Neutron Transport} 
\label{sec:transpssort}
Following the formalism used for the transport equation in \citet{brantley2000simplified}, consider a domain $V$ consisting of $I$ simply connected nonoverlapping homogeneous material regions $\left(V_i\right)_{i=1}^I$ with outer boundary
\begin{gather}
\partial V = \partial  \bigcup_{i=1}^{I} V_i
\end{gather}
and material interfaces
\begin{gather}
\partial V_{ij} = V_i \cap V_j.
\end{gather}
An outward-facing normal vector from material $V_i$ that is defined on $\partial V_i$ is represented with $\hat{\bm{n}}_i$.
With this, the fixed-source steady-state neutron transport equation can then be expressed as
\begin{gather}
\hat{\bm{\Omega}} \! \cdot \! \nabla \psi_{i}^{(n)}(\bm{r}, \hat{\bm{\Omega}}, E) + \Sigma_{t,i}^{(n)}(E) \psi^{(n)}_{i}(\bm{r}, \hat{\bm{\Omega}}, E) \\ = \int_0^\infty\int_{4\pi}\Sigma_{s, i}^{(n)}(\hat{\bm{\Omega}}'\cdot \hat{\bm{\Omega}}, E' \rightarrow E) \ \psi_{i}^{(n)}(\bm{r}, \hat{\bm{\Omega}}', E') d\Omega'dE' + \frac{Q_{i}^{(n)}(\bm{r}, E)}{4\pi}, \quad \bm{r} \in V_i.
\end{gather}
In this expression,  $\psi_{i}^{(n)}(\bm{r}, \hat{\bm{\Omega}}, E)$ is the angular neutron flux in region $i$.
It varies over space ($\bm{r}$), direction of travel ($\hat{\bm{\Omega}}$) and energy ($E$).
The $^{(n)}$ flag is not an exponent but a designation that this quantity corresponds to the neutron flux.
Then, $\Sigma_{t,i}^{(n)}(E)$ is the total cross section in the $i$-th region.
The differential scattering cross section for region $i$ from energy $E'$ to $E$ with scattering angle cosine $\hat{\bm{\Omega}}'\cdot \hat{\bm{\Omega}}$ is represented with $\Sigma_{s, i}^{(n)}(\hat{\bm{\Omega}}'\cdot \hat{\bm{\Omega}}, E' \rightarrow E)$.
Finally, $Q_{i}^{(n)}(\bm{r}, E)$ is a spatially distributed volumetric isotropic neutron source.

At material interfaces, 
\begin{gather}
\label{eq:transportICs}
\psi^{(n)}_i(\bm{r}, \hat{\bm{\Omega}}, E) =  \psi^{(n)}_j(\bm{r}, \hat{\bm{\Omega}}, E), \quad \bm{r} \in \partial V_{ij},
\end{gather}
and at the external boundaries
\begin{gather}
\psi^{(n)}_i(\bm{r}, \hat{\bm{\Omega}}, E) = 0, \quad \bm{r} \in \partial V_{i}, \quad\hat{\bm{n}}_i\cdot \hat{\bm{\Omega}}<0.
\end{gather}
No boundary sources are considered in this work.
It is also helpful to define the neutron scalar flux as
\begin{gather}
\phi^{(n)}_i(\bm{r}, E) = \int_{4 \pi}\psi^{(n)}_i(\bm{r}, \hat{\bm{\Omega}}, E) d\Omega.
\end{gather}

\subsection{Transport Equation for Approximate Heat Conduction}
\label{sec:transeqapprox}
Now, consider a heat conduction problem imposed on the same regions as the neutron transport problem.
The equilibrium temperature distribution within a material region $T_i(\bm{r})$ can be described by the steady-state conduction equation,
\begin{gather}
\nabla^2 T_i(\bm{r}) + \frac{Q^{(h)}(\bm{r})}{k_i} = 0, \quad \bm{r} \in V_i.
\end{gather}
Here, $k_i$ represents the thermal conductivity in $V_i$ and $Q^{(h)}(\bm{r})$ represents a volumetric heat generation rate in $V_i$.
For this work, $Q^{(h)}(\bm{r})$ is restricted to have the same spatial distribution as $Q_{i}^{(n)}(\bm{r}, E)$.
Mathematically, 
\begin{gather}
\frac{Q^{(h)}_i(\bm r)}
{\displaystyle \sum_{j}\int_{V_j} Q^{(h)}_j(\bm r')\,d \bm{r}'}
\;=\;
\frac{\displaystyle \int_{0}^{\infty} Q^{(n)}_{i}(\bm r,E)\,dE}
{\displaystyle \sum_{j}\int_{V_j}\!\int_{0}^{\infty} Q^{(n)}_{j}(\bm r',E)\,dE\,d \bm{r}'},
\qquad \forall\,\bm r\in V_i,\ \forall i .
\end{gather}
Also, no heat sources are considered on material interfaces or boundaries.
As noted in \citet{Jun2007MonteCarloHeat} and relevant to the application of the scaling factor to the mean free path as presented in Section \ref{sec:gea},  scaling $k_i$ and $Q^{(h)}_i$ by the same factor does not change the solution to this differential equation.
Then, inhomogeneous Robin boundary conditions can be specified with
\begin{gather}
A_i  \hat{\bm{n}}_i \cdot \nabla T_i(\bm{r})  + B_i T_i(\bm{r}) = C_i(\bm{r}), \quad \bm{r} \in \partial V_i.
\end{gather}
Here, $A_i$, $B_i$ and $C_i(\bm{r})$ are problem-dependent.
For practical problems, restrict $A_i\ge 0$ and $B_i\ge 0$.
Of course, this specification allows for the treatment of both Dirichlet and Neumann boundary conditions by setting $A$ or $B$ to 0, respectively. 
On material region interfaces, continuity of heat flux and temperature are enforced with
\begin{gather}
T_i(\bm{r})=T_j(\bm{r}), \quad \bm{r} \in \partial V_{ij},
\end{gather}
and
\begin{gather}
k_i \hat{\bm{n}}_i \cdot \nabla T_i(\bm{r})  = -k_j \hat{\bm{n}}_j \cdot \nabla T_j(\bm{r}) , \quad \bm{r} \in \partial V_{ij}.
\end{gather}

The following section will explicitly discuss boundary condition homogenization.
This is a necessary step for compatibility with this initial presentation of MOSS.
Then, Section \ref{sec:gea} will present the transport equation that approximates the conduction equation.
Finally, Section \ref{sec:acaa} will present the corresponding boundary conditions.

\subsubsection{Boundary Condition Homogenization}
\label{sec:bch}
In order to apply the approximations discussed in the next two sections, the boundary conditions of the problem must be homogenized.
Then, only the source-driven component of the temperature distribution with homogeneous boundary conditions will be treated with the transport approximation.
The boundary-driven harmonic component of the temperature distribution with inhomogeneous boundary conditions will need to be solved separately using deterministic methods.
Although the content of this subsection should be considered fairly standard, it is included for completeness.

First, express the temperature distribution in each material region as a summation of two components
\begin{gather}
T_i(\bm{r}) = \tilde{T}_i(\bm{r}) + \breve{T}_i(\bm{r}), \quad \bm{r} \in V_i,
\end{gather}
where $\tilde{T}_i(\bm{r})$ \hl{represents the source-driven component of the temperature distribution with homogeneous boundary conditions that captures the impact of the distributed heat source while} $\breve{T}_i(\bm{r})$ \hl{is selected to be a boundary-driven harmonic function subject to the original inhomogeneous boundary conditions.}

Being a harmonic function, Laplace's equation governs the behavior of $\breve{T}_i(\bm{r})$,
\begin{gather}
\nabla^2\breve{T}_i(\bm{r})=0, \quad \bm{r} \in V_i,
\end{gather}
and is subject to
\begin{gather}
A_i \hat{\bm{n}}_i \cdot \nabla \breve{T}_i(\bm{r})  + B_i \breve{T}_i(\bm{r}) = C_i(\bm{r}), \quad \bm{r} \in \partial V_i, \\
\breve{T}_i(\bm{r})=\breve{T}_j(\bm{r}), \quad \bm{r} \in \partial V_{ij}, \\
k_i \hat{\bm{n}}_i \cdot\nabla \breve{T}_i(\bm{r}) = -k_j \hat{\bm{n}}_j \cdot\nabla \breve{T}_j(\bm{r}), \quad \bm{r} \in \partial V_{ij}.
\end{gather}

\hl{With this specification, the governing equation and auxiliary conditions for the source-driven component} $\tilde{T}_i(\bm{r})$ are
\begin{gather}
\nabla^2 \tilde{T}_i(\bm{r}) + \frac{Q^{(h)}(\bm{r})}{k_i} = 0, \quad \bm{r} \in V_i, \\
A_i \nabla \tilde{T}_i(\bm{r}) \cdot \hat{\bm{n}}_i + B_i \tilde{T}_i(\bm{r}) = 0, \quad \bm{r} \in \partial V_i, \label{eq:tildebc}\\
\tilde{T}_i(\bm{r})=\tilde{T}_j(\bm{r}), \quad \bm{r} \in \partial V_{ij}, \label{eq:Tcont} \\
k_i \hat{\bm{n}}_i \cdot \nabla \tilde{T}_i(\bm{r})  = -k_j \hat{\bm{n}}_j \cdot \nabla \tilde{T}_j(\bm{r}) , \quad \bm{r} \in \partial V_{ij} \label{eq:fluxcont}.
\end{gather}
This is the form treated in the following subsections.

\subsubsection{Governing Equation Approximation}
\label{sec:gea}
The angular flux of the theoretical heat particle will be expressed as $\psi^{(h)}(\bm{r}, \hat{\bm{\Omega}})$ with spatial and angular dependence.
The $^{(h)}$ superscript indicates that this quantity corresponds to the theoretical heat particle. 
Adapted from \citet{Prinja2010}, the source-driven steady-state Boltzmann transport equation for a purely scattering medium can be expressed as
\begin{gather}
\hat{\bm{\Omega}} \! \cdot \! \nabla \psi_i^{(h)}(\bm{r}, \hat{\bm{\Omega}}) + \Sigma^{(h)}_{s,i}\psi^{(h)}_i(\bm{r}, \hat{\bm{\Omega}}) = \frac{\Sigma_{s,i}^{(h)}}{4\pi}\int_{4\pi} \psi_i^{(h)}(\bm{r}, \hat{\bm{\Omega}}')\, d\Omega' + \frac{Q_i^{(h)}(\bm{r})}{4\pi}, \quad \bm{r} \in V_i.
\end{gather}
Using conventional diffusion/transport relationships for purely isotropic scattering media the scattering cross section could be selected as $\frac{1}{3k_i}$ \citep{fraley1980monte}.
However, using the definition of $\beta$ presented in \citet{fraley1980monte}, selecting
\begin{gather}
\Sigma^{(h)}_{s,i} = \frac{1}{3k_i \beta}, \quad \forall i
\end{gather}
preserves the diffusion solution while changing the transport process to be more ``diffusion-like''.
Lower values of this scaling factor, $\beta$ , will lead to better estimations.
Then, the space-dependent temperature can be expressed as
\begin{gather}
\tilde{T}_i(\bm{r}) \approx \beta \int_{4\pi} \psi^{(h)}_i(\bm{r}, \hat{\bm{\Omega}}') d\Omega' , \quad \bm{r} \in V_i.
\end{gather}
The approximate relation is used to acknowledge the approximation introduced with $\psi^{(h)}(\hat{\bm{\Omega}}, \bm{r})$ being determined by a transport process.

\subsubsection{Auxiliary Conditions Approximation}
\label{sec:acaa}
For the multiregion problems under consideration, auxiliary conditions include both the interface and boundary conditions.
In particular, obtaining auxiliary conditions for the analogous transport problem is nontrivial.
The approaches used in this work rely on an inexact relationship between a transport-governed partial heat particle currents and the temperature.
Namely, with an angular linearization to $\psi^{(h)}(\hat{\bm{\Omega}}, \bm{r})$ and Fick's law approximation for the total particle current, 
\begin{align}
J^{(h)+}_i(\bm{r}) &\approx \frac{1}{4}\,\tilde{T}_i(\bm{r}) - \frac{1}{2}\beta k_i \hat{\bm{n}}_i\!\cdot\!\nabla \tilde{T}_i(\bm{r}), \quad \bm{r} \in \partial V_i, \label{eq:Jp_aprox} \\
J^{(h)-}_i(\bm{r}) &\approx \frac{1}{4}\,\tilde{T}_i(\bm{r}) + \frac{1}{2}\beta k_i \hat{\bm{n}}_i\!\cdot\!\nabla \tilde{T}_i(\bm{r}), \quad \bm{r} \in \partial V_i \label{eq:Jm_aprox}.
\end{align}
Here, 
\begin{gather}
J^{(h)+}_i(\bm{r}) \equiv \int_{\hat{\bm{\Omega}}\cdot \hat{\bm{n}}_i>0} \left(\hat{\bm{\Omega}}\cdot \hat{\bm{n}}_i\right) \psi_i^{(h)}(\bm{r}, \hat{\bm{\Omega}}) d\Omega, \\
J^{(h)-}_i(\bm{r}) \equiv \int_{\hat{\bm{\Omega}}\cdot \hat{\bm{n}}_i<0} \lvert\hat{\bm{\Omega}}\cdot \hat{\bm{n}}_i\rvert \, \psi_i^{(h)}(\bm{r}, \hat{\bm{\Omega}}) d\Omega.
\end{gather}
Physically, $J^{(h)+}(\bm{r})$ and $J^{(h)-}(\bm{r})$ represent outward and inward particle flow rates across $\partial V_i$ at $\bm{r}$ and are often referred to as partial currents.

Combining Equations \ref{eq:Jp_aprox} and \ref{eq:Jm_aprox} with Equations \ref{eq:Tcont} and \ref{eq:fluxcont} yields a current-conservation interface condition
\begin{gather}
J_i^{(h)+}(\bm{r})=J_j^{(h)-}(\bm{r}), \quad J_i^{(h)-}(\bm{r})=J_j^{(h)+}(\bm{r}), \quad \bm{r} \in \partial V_{ij}.
\end{gather}
This interface condition is necessary for the flux-continuity interface conditions provided in Equation \ref{eq:transportICs} to be satisfied.
Therefore, the flux-continuity will be applied to both the neutron and heat particle transport problem identically.
This is a straightforward method for ensuring the interface conditions associated with the heat conduction problem are satisfied.

Now, the boundary conditions given in Equation \ref{eq:tildebc} are treated.
In \citet{Jun2007MonteCarloHeat}, a geometric boundary correction is applied to mimic the typical extrapolated boundary condition used in neutronics problems.
Although this method is effective in many problems, \citet{fraley1980monte} points out that ``this type of correction is not always practical for problems with complex shapes or with several different types 
of materials on the external boundaries.''
An albedo-based approach is then presented which will be used in the current work.
This albedo-based approach can lead to significantly higher computation times but is not subject to the geometric limitations associated with the approach demonstrated in \citet{Jun2007MonteCarloHeat}.

First, define some $H_i$ such that
\begin{gather}
H_i = \begin{cases}
k_i\frac{B_i}{A_i} & A_i > 0 \\
\infty & A_i = 0
\end{cases}.
\end{gather}
This corresponds to the definition of the conventional heat transfer coefficient.
Then, define some surface reflection probability, or albedo, $\alpha^{(h)}_i$, as
\begin{gather}
\alpha^{(h)}_i = \frac{J_i^{(h)-}(\bm{r})}{J_i^{(h)+}(\bm{r})}.
\end{gather}
Using Equations \ref{eq:Jp_aprox} and \ref{eq:Jm_aprox} and the boundary conditions given in Equation \ref{eq:tildebc}, an expression for the albedo is
\begin{gather}
\alpha^{(h)}_i = \frac{1 - 2\beta H_i}{1 + 2\beta H_i}, \quad \bm{r} \in \partial V_i.
\end{gather}
As $\alpha^{(h)}_i$ can be less than 0, this boundary condition can be simulated in a Monte Carlo code by multiplying the weight of the particle by $\text{sign}\left(\alpha^{(h)}_i\right)$ and reflecting the particle with probability $\lvert \alpha^{(h)}_i \rvert$.
As noted in \citet{fraley1980monte}, problems with exclusively Dirichlet and/or Neumann boundary conditions will be nonconvergent due to the absence of any way to end the particle history.
The recommendation was given to, instead, set $H_i$ on the Dirichlet boundaries to be high but not infinite.
This will incur additional error but yield completable calculations.
Future work may explore the use of rouletting or splitting \citep{gonzalez2021choosing} to mitigate this issue.

\hl{The correspondence between the transport and conduction auxiliary conditions is used here in a formal diffusion-limit sense.
In small mean-free-path limits, transport solutions may develop boundary or interface layers where the angular flux adjusts rapidly to imposed kinetic boundary or interface conditions} \citep{bensoussan1979boundary}.
\hl{A rigorous layer analysis for the albedo boundary conditions and material-interface treatment used in MOSS is outside the scope of this initial presentation.
The numerical examples considered here did not exhibit visible layer-induced artifacts at the resolution shown, but such effects may occur for other choices of boundary conditions, material contrasts, or values of} $\beta$.

\subsection{Simultaneous Solutions with Correlated Sampling}
\label{sec:sscorsm}
With the linearity of the transport equation, two normalized fluxes can be introduced
\begin{gather}
\ddot{\psi}^{(n)}_i(\bm{r}, \hat{\bm{\Omega}}, E) = \frac{\psi^{(n)}_i(\bm{r}, \hat{\bm{\Omega}}, E)}{\sum_i^I \int_0^\infty \int_{V_i} Q_i^{(n)}(\bm{r}',E') d\bm{r}'dE'}, \quad \bm{r} \in V_i,
\end{gather}
and
\begin{gather}
\ddot{\psi}^{(h)}_i(\bm{r}, \hat{\bm{\Omega}}) = \frac{\psi^{(h)}_i(\bm{r})}{\sum_i^I  \int_{V_i} Q_i^{(h)}(\bm{r}') d\bm{r}'}, \quad \bm{r} \in V_i.
\end{gather}
These normalized fluxes are solutions to the aforementioned transport equations with unit-strength sources.

Thorough discussions on the derivation of correlated sampling method are given in \citet{majumdar1996development} and \citet{rief1984generalized}.
For this work, derivations will not be presented.
The difference in the normalized angular fluxes can be written as
\begin{equation}
\begin{gathered}
\label{eq:corrsampdiff}
\ddot{\psi}^{(h)}(\bm{r}, \hat{\bm{\Omega}}) - \int_0^\infty\ddot{\psi}^{(n)}(\bm{r}, \hat{\bm{\Omega}}, E) dE
 = \\ \int_0^\infty\sum_{s=1}^\infty \int_R \int_R \cdots \int_R \left( \prod_{\ell=0}^s F_\ell - 1 \right) \left(\prod_{\ell=1}^s\kappa^{(n)}\left( \bm{u}_{\ell+1}, \bm{u}_\ell \right) d\bm{u}_\ell \, G^{(n)}(\bm{u}_1, \bm{u}_0)\frac{\ddot{Q}(\bm{u}_0)}{4\pi}\right) d\bm{u}_0 dE .
 \end{gathered}
\end{equation}
In this equation, a new variable is introduced: $\bm{u}_\ell = \left[ \bm{r}_\ell, \hat{\bm{\Omega}}_\ell, E_\ell\right]$.
This variable is a notational convenience used to represent a complete description of the state of the particle after $\ell$ collisions.
Then, $\int_R$ denotes integration over the full particle phase space $R$, that is, all admissible states of $\bm{u}_\ell$.
To streamline notation, the region indices $i$  and  explicit sums over the $I$ regions are omitted.
All region-dependent quantities can be interpreted as piecewise functions of $\bm r$ that equal their region-specific values on $V_i$.
Also,
\begin{gather}
\ddot{Q}(\bm{r}, E) \equiv \frac{Q^{(n)}(\bm{r}, E)}{\int_0^\infty \int_V Q^{(n)}(\bm{r}', E') d\bm{r}' dE'}.
\end{gather}

The discussion on the remaining quantities in Equation \ref{eq:corrsampdiff} involve  transport kernels. 
To begin, consider the transport kernel for the neutron, 
\begin{gather}
\label{eq:transportker}
\kappa^{(n)}\left( \bm{u}_{\ell+1}, \bm{u}_\ell \right) = C^{(n)}( \hat{\bm{\Omega}}_{\ell+1}, E_{\ell+1}, \bm{r}_\ell, \hat{\bm{\Omega}}_\ell, E_\ell)G^{(n)}(\bm{r}_{\ell+1},\bm{r}_{\ell}, E_{\ell}).
\end{gather}
This presentation follows the notation in \citet{majumdar1996development} and \citet{rief1984generalized} where $\bm{u}_\ell$ and its constituents $\bm{r}_\ell$, $\hat{\bm{\Omega}}_\ell$ and $E_\ell$ are presented alongside each other in the same equation.
Here, $\kappa^{(n)}\left( \bm{u}_{\ell+1}, \bm{u}_\ell \right)$ is the transport kernel from state $\bm{u}_\ell$ to $\bm{u}_{\ell+1}$ for the neutron.
It is factored into a product of the neutron collision kernel ($ C^{(n)}$) and neutron translation kernel ($G^{(n)}$).
The collision kernel is further subdivided into collision types
\begin{gather}
C^{(n)}( \hat{\bm{\Omega}}_{\ell+1}, E_{\ell+1}, \bm{r}_\ell, \hat{\bm{\Omega}}_\ell, E_\ell) = \sum_k p_k \,C^{(n)}_k( \hat{\bm{\Omega}}_{\ell+1}, E_{\ell+1}, \bm{r}_\ell, \hat{\bm{\Omega}}_\ell, E_\ell).
\end{gather}
Here, $p_k$ is the probability of a collision of type $k$ and $C^{(n)}_k$ is the corresponding collision kernel. 
Expressions for $C^{(n)}_k$ vary by reaction type but are well-established.
Then, the translation kernel for free-flight in a homogeneous material region can be expressed with
\begin{gather}
G^{(n)}(\bm{r}_{\ell+1},\bm{r}_{\ell}, E_{\ell}) = \Sigma^{(n)}_t(\bm{r}_\ell, E_\ell) \exp{\left(-\Sigma_t^{(n)}(\bm{r}_{\ell}, E_\ell) ||\bm{r}_{\ell+1} - \bm{r}_\ell||_2\right)}.
\end{gather}
Treating translations across material interfaces is not explicitly discussed but, to summarize, can be treated using $\kappa^{(n)}\left( \bm{u}_{\ell+1}, \bm{u}_\ell \right) = \exp{\left(-\Sigma_t^{(n)}(\bm{r}_{\ell}, E_\ell) ||\bm{r}_{\ell+1} - \bm{r}_\ell||_2\right)}$  and advancing the $\ell$ index.

Moving on, consider the transport kernel for the heat particle, $\kappa^{(h)}$, 
\begin{gather}
\label{eq:transportkerheat}
\kappa^{(h)}\left( \bm{u}_{\ell+1}, \bm{u}_\ell \right) = C^{(h)}( \bm{r}_\ell)G^{(h)}(\bm{r}_{\ell+1},\bm{r}_{\ell}).
\end{gather}
As there is only one reaction type (isotropic scatter), the collision kernel can be expressed with
\begin{gather}
 C^{(h)}(  \bm{r}_\ell) = \frac{\Sigma_s^{(h)}(\bm{r}_\ell)}{4\pi}.
\end{gather}
Then, the translation kernel for free-flight in a homogeneous material region can be expressed with
\begin{gather}
G^{(h)}(\bm{r}_{\ell+1},\bm{r}_{\ell}) = \Sigma^{(h)}_s(\bm{r}_\ell) \exp{\left(-\Sigma_s^{(h)}(\bm{r}_{\ell}) ||\bm{r}_{\ell+1} - \bm{r}_\ell||_2\right)}.
\end{gather}

The final variable that requires discussion from Equation \ref{eq:corrsampdiff} is $F_\ell$.
This is the weighting factor for the heat transfer problem, it is the ratio of the two transport kernels,
\begin{gather}
F_\ell = \begin{cases}
\frac{ G^{(h)}(\bm{u}_1, \bm{u}_0)}{ G^{(n)}(\bm{u}_1, \bm{u}_0)} & \ell = 0 \\
\frac{\kappa^{(h)}(\bm{u}_{\ell+1}, \bm{u}_{\ell})}{\kappa^{(n)}(\bm{u}_{\ell+1}, \bm{u}_{\ell})} & \ell > 0\\
\end{cases}.
\end{gather}
The disregard for the collision kernel in $\ell=0$ arises to carry the particle from its birth to $\bm{u}_1$ where the first collision will have occurred.

\subsection{Branching for Boundary Condition Satisfaction}
\label{sec:branching}
Unfortunately, some portion of the histories in the heat transport problem cannot occur in the neutron transport problem.
As discussed in Section \ref{sec:bch}, heat particles may reflect at some boundaries.
This behavior has no analogue in neutron transport.
Mathematically, imposing this reflection on the neutron problem can be understood as including an action where $\kappa^{(n)} = 0$ and $F_\ell$ becomes undefined.
Therefore, a splitting-based technique is used.
Specifically, upon intersection of a particle path with an outer boundary, the heat particle and neutron are separated.
The particle history will continue with probability $\alpha^{(h)}_i$ but no neutron-related quantities will be tallied.
This leads to computational time dedicated solely to tracking the heat particle.

\subsection{Violations of the Central Limit Theorem}
\label{sec:clt_vio}
Another significant limitation in the MOSS methodology is the possibility for the variance associated with heat particle-related estimators to become unbounded.
This arises because the central limit theorem, which would ordinarily guarantee convergence of these estimators as the sample size increases, is inapplicable when the variance of the observations within each estimator is not finite \citep{billingsley1995probability}.
This can occur if the stochastic behavior of the heat particle and neutron are sufficiently dissimilar.
More detailed discussion on this in the context of conventional correlated sampling is given in \citet{rief1984generalized} and \citet{Dejonghe1981-MonteCarlo}.
\ref{ap:clt_vio} \hl{provides a numerical illustration of this behavior for the slab demonstration problem by varying the Material A thermal conductivity while holding the remaining problem definition fixed.}

In the results presented in Section \ref{sec:resultss}, the random walk procedure is modified to only use the scattering properties of the neutronic process.
Then, between scatter reactions, there is a chance for an absorption reaction to occur.
If absorption does occur, no more neutron-related estimators are accumulated for the remainder of the walk.
This leads to more computational time dedicated solely to tracking the heat particle.
Future work should develop better methods for addressing this problem.

\section{Results}
\label{sec:resultss}
Two demonstration problems are presented to demonstrate MOSS.
The first problem consists of a 2-region problem with slab geometry.
For this problem, the solution for the \hl{boundary-driven component} of the temperature distribution is the null function.
For the neutron transport solution, the results from MOSS are compared with those calculated with OpenMC \citep{romano2015openmc}.
Then, for the temperature solution, the results from MOSS are compared to both an analytical solution and those calculated with OpenMC applied to the heat particle transport problem.
The comparison with OpenMC is done to isolate the use of the correlated sampling method where the transport approximation to the conduction problem will impact both solution methods equally.
The second problem is built to replicate a 2D pin cell of a graphite-moderated heat pipe reactor.
OpenMC is again used for providing a non-simultaneous solution to the neutron transport problem.
Then, the MOOSE heat transfer module \citet{harbour2025moose} is used to supply the boundary-driven harmonic component of the temperature distribution, $\breve{T}$, as well as the full temperature distribution for comparison with MOSS.
The neutronics component of these demonstration problems will use multigroup transport for the neutronics problem.
Conceptually, this has no significant impact on the results presented \hl{as the method is equally applicable to continuous energy problems}.

Both demonstration problems use the same two materials, whose properties are given in Table \ref{tab:xsecs}.
In this table, the total neutronic cross section for the $g$-th energy group is given with $\Sigma_{t,g}^{(n)}$.
Then, the total scattering cross section from group $g'$ to $g$ and its first moment are given by $\Sigma_{s0, g'\rightarrow g}^{(n)}$ and $\Sigma_{s1, g'\rightarrow g}^{(n)}$, respectively.
The thermal conductivity is given for each material in this table as well as $\beta$ used for the problem.
Also included in this table is the total scattering cross section of the heat particle, $\Sigma_s^{(h)}$, calculated from $k$ and $\beta$.
For this preliminary presentation of MOSS, these material properties are carefully selected such that the neutronic and heat particle stochastic processes are similar.
This selection is intended to avoid the variance growth discussed in Section \ref{sec:clt_vio}.
\ref{ap:clt_vio} \hl{illustrates the sensitivity of the slab demonstration problem to this selection by varying the Material A thermal conductivity.}

\begin{table}[htbp]
\centering
\caption{Properties of materials used in demonstration problems. All cross sections given in cm$^{-1}$. Thermal conductivity is given in W/cm-K.}
\label{tab:xsecs}
\begin{tabular}{|cc|c|cc|cc|c|c|c|}
\hline
\multicolumn{2}{|c|}{\multirow{2}{*}{}}                   & \multirow{2}{*}{$\Sigma_{t,g}^{(n)}$} & \multicolumn{2}{c|}{$\Sigma_{s0, g'\rightarrow g}^{(n)}$} & \multicolumn{2}{c|}{$\Sigma_{s1, g'\rightarrow g}^{(n)}$} & \multirow{2}{*}{$k$} & \multirow{2}{*}{$\Sigma_s^{(h)}$} & \multirow{2}{*}{$\beta$} \\ \cline{4-7}
\multicolumn{2}{|c|}{}                                    &                                   & \multicolumn{1}{c|}{$g'=1$}            & $g'=2$            & \multicolumn{1}{c|}{$g'=1$}            & $g'=2$            &                      &                                   &                          \\ \hline
\multicolumn{1}{|c|}{\multirow{2}{*}{Material A}} & $g=1$ & 0.535                             & \multicolumn{1}{c|}{0.495}             & 0                 & \multicolumn{1}{c|}{0.030}             & 0                 & \multirow{2}{*}{6.4} & \multirow{2}{*}{0.521}            & \multirow{4}{*}{0.1}     \\ \cline{2-7}
\multicolumn{1}{|c|}{}                            & $g=2$ & 0.650                             & \multicolumn{1}{c|}{0.020}             & 0.515             & \multicolumn{1}{c|}{0.003}             & 0.002             &                      &                                   &                          \\ \cline{1-9}
\multicolumn{1}{|c|}{\multirow{2}{*}{Material B}} & $g=1$ & 1.050                             & \multicolumn{1}{c|}{0.350}             & 0                 & \multicolumn{1}{c|}{0.045}             & 0                 & \multirow{2}{*}{3.21} & \multirow{2}{*}{1.038}           &                          \\ \cline{2-7}
\multicolumn{1}{|c|}{}                            & $g=2$ & 1.045                             & \multicolumn{1}{c|}{0.680}             & 1.035             & \multicolumn{1}{c|}{0.060}             & 0.003             &                      &                                   &                          \\ \hline
\end{tabular}
\end{table}

\subsection{Slab Demonstration Problem}
\begin{figure}[htp]
    \centering
    \includegraphics[width=\textwidth]{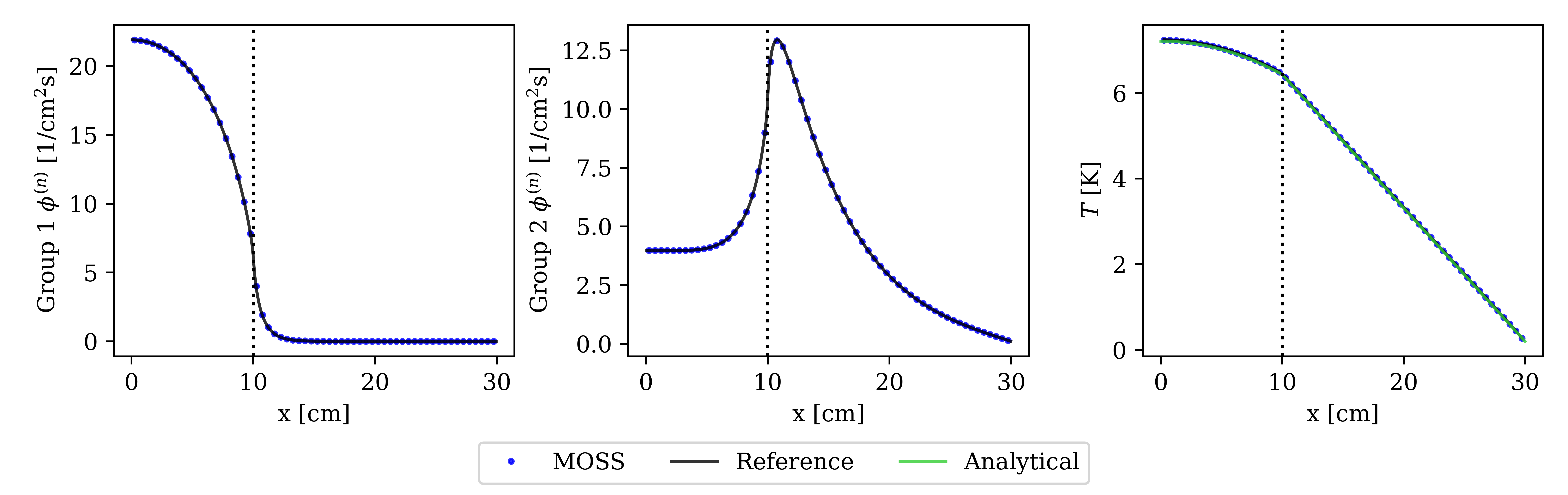}
    \caption{Results from MOSS, OpenMC (reference) and analytical solutions for slab geometry demonstration problem. A vertical line is placed at 10 cm to indicate the division between the two material regions.}
    \label{fig:ref}
\end{figure}

The first demonstration problem is only heterogeneous along the x-axis, creating a two region slab geometry with Material A spanning from 0 cm to 10 cm ($i=1$ in this region) and Material B extending from 10 cm to 30 cm ($i=2$ in this region).
\hl{The boundary-driven harmonic component of the temperature distribution, }$\breve{T}_i(\bm r)$, \hl{is globally zero for this configuration, so this slab problem shows the ability of MOSS to approximate} $T_i(\bm r)$ \hl{without the additional complication of boundary-condition homogenization.}
Within Material A, a spatially-uniform neutron source emits 1 $\frac{\text{neutrons}}{\text{cm}^3\text{s}}$ and 0.1 $\frac{\text{neutrons}}{\text{cm}^3\text{s}}$ in the energy group 1 and 2, respectively.
The spatially-uniform heat source has a strength of 0.1 $\frac{\text{W}}{\text{cm}^3}$.
Practically, this source could represent a spontaneous fission neutron source whose reactions release both neutrons and heat.
At the left boundary, a reflective boundary condition is imposed for both the heat conduction and transport problems.
Then, at the right boundary, the boundary conditions given in Equation \ref{eq:transportICs} are imposed for the neutron transport problem.
Figure \ref{fig:ref} shows the group-wise neutron scalar fluxes and temperature distribution calculated with $B_2/A_2 = 1/(2\beta k_2)$ imposed on the right boundary for the heat conduction problem.
This boundary condition leads to $\alpha^{(h)}_2=0$ and eliminates the need for branching at the boundary.
For each quantity, OpenMC is applied for the heat conduction and neutron transport problems independently to generate a reference result.
For all Monte Carlo-based solution methods, a sufficient number of histories were run to ensure convergence at the scale apparent in the provided plot.
\hl{A separate convergence verification for this slab problem is provided in} \ref{ap:cv}.
The analytical solution to the heat conduction problem is also provided.
The neutron fluxes from MOSS and the OpenMC calculation show close agreement, demonstrating that correlated sampling has no impact on the base neutron transport calculation.
For the temperature distribution, both OpenMC and MOSS show the same slight overestimation, \hl{about 0.016 K absolute error on average}, compared to the analytic solution within the source region.
\hl{This minor disagreement is a result of applying the transport approximation to the diffusion equation and is explored in more detail in} \ref{ap:optthic}.
\hl{Previous studies} \citep{Fraley1977MonteCarlo, song2007improved} \hl{have explored this approximation with more depth.}

As discussed in Section \ref{sec:branching}, some portions of the particle history do not tally neutron-related quantities, there is some computational time dedicated solely to solving the conduction equation.
For the demonstration problem shown in Figure \ref{fig:ref}, this occurs only when the neutron is absorbed.
However, with any other boundary conditions ($\alpha^{(h)}_2 \ne 0$), this also occurs any time the history encounters an outer boundary and the heat particle requires the history to continue after a reflection.
To illustrate the need for adjustments to the random walk procedure that will more carefully balance variance reduction and simultaneous solution time, Table \ref{tab:comptime_res} provides some insight on how much computational time is dedicated solely to solving the heat conduction equation for this demonstration problem.
The column ``Conduction Only'' in this table gives the percentage of steps in all random walks that do not provide tallies to estimators related to neutronic quantities.
These steps provide tallies only for the heat conduction problem.
From these results, the case of $\alpha^{(h)}_2 = 0$ leads to absorption being the only cause for particle branching.
In this case,  only 1.8\% of the computational time is committed to simultaneously solving both equations.
The further imposition of heat-reflective boundary conditions ($\alpha^{(h)}_2 > 0$) worsens this problem.
Although this computational inefficiency is highly problem-dependent, some future adjustments to the random walk procedure could potentially address this concern.
For example, including an adjustable particle weight corresponding to the neutronic process would enable more flexibility in the random walk provided the neutronic weight is updated accordingly.
Separately, to minimize the impact of boundary reflections, the heat particle survival probability at the boundary can be reduced with the heat particle weight increased proportionally.

\begin{table}[htb!]
\centering
\caption{Percentage of computational effort dedicated to solving heat conduction only. Here, $\rho$ is a dividing factor applied to the boundary condition such that $B_2/A_2 = 1/(2\beta k_2 \rho)$. This parameter is increased to change the boundary condition and make reflection of the heat particle more likely.}
\label{tab:comptime_res}
\begin{tabular}{ccc}
\hline
$\rho$ & $\alpha^{(h)}_2$ & Conduction Only \\ \hline
1      & 0          & 98.2 \%       \\
5      & 0.67       & 98.5 \%       \\
10     & 0.82       & 98.7 \%       \\
15     & 0.88       & 98.9 \%       \\
20     & 0.90       & 99.1 \%       \\ \hline
\end{tabular}
\end{table}

\subsection{Pin Cell Demonstration Problem}
\begin{figure}[htp]
    \centering
    \includegraphics[width=.5\textwidth]{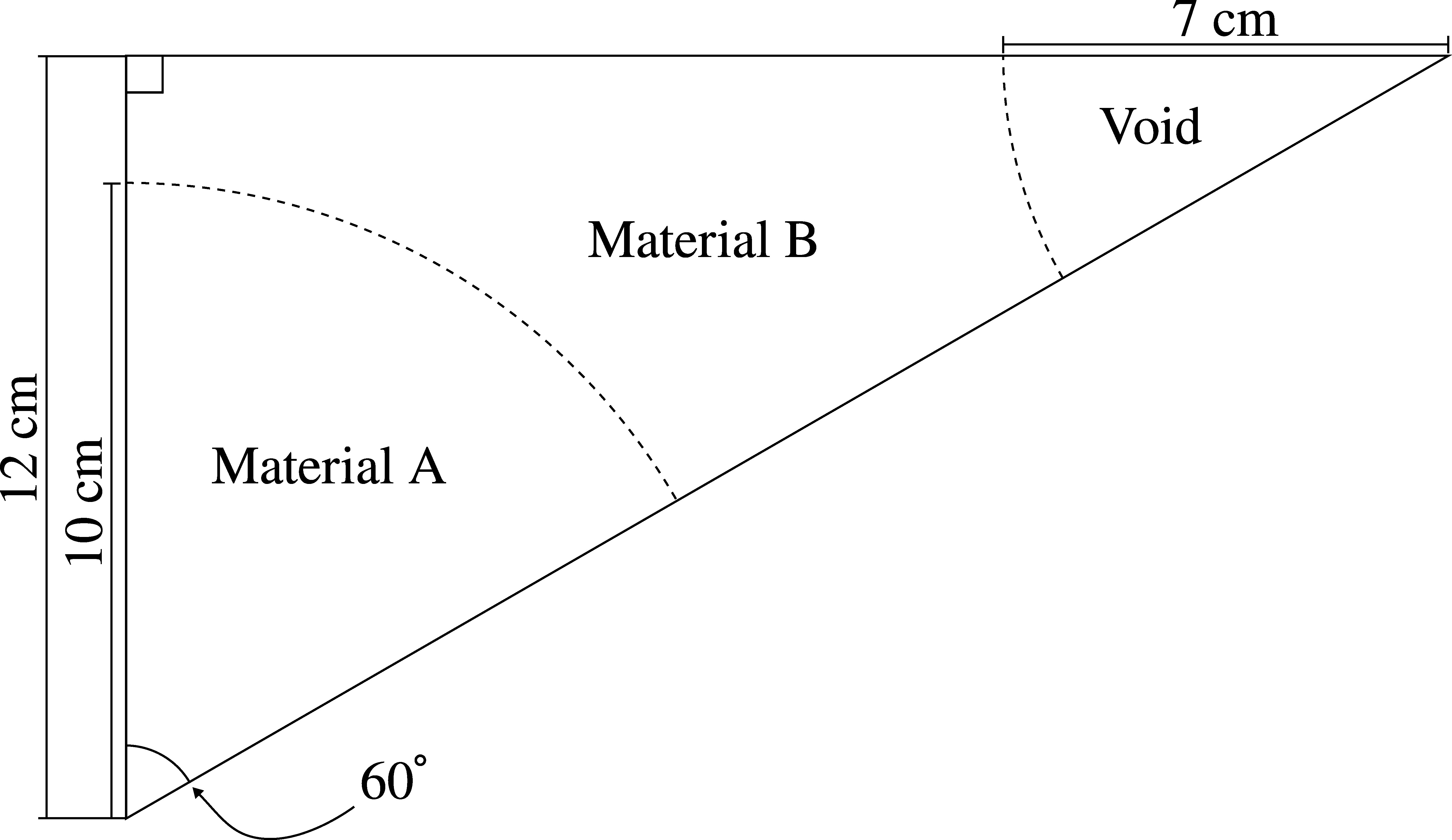}
    \caption{Dimensioned geometry for pin cell demonstration problem with voided region.}
    \label{fig:2ddiag}
\end{figure}

This demonstration problem is created to demonstrate the use of MOSS in a more realistic 2D reactor geometry.
The geometry shown in Figure \ref{fig:2ddiag} consists of a triangular fuel region (Material A), a surrounding moderator (Material B) and a peripheral void region representing the vapor core of a heat pipe.
To avoid introducing different cross sections from slab demonstration problem, the entire geometry is enlarged to ensure an adequate optical thickness of each region.
This is done to strengthen the transport approximation to the conduction equation for demonstration.
The region filled with Material A is, again, the source region emitting 1 $\frac{\text{neutrons}}{\text{cm}^3\text{s}}$ and 0.1 $\frac{\text{neutrons}}{\text{cm}^3\text{s}}$ in the energy group 1 and 2, respectively.
This region also has a heat source with strength of 20 $\frac{\text{W}}{\text{cm}^3}$.
Both of these sources are spatially uniform.
A region filled with Material B surrounds the source region.
Finally, a heat pipe region is included in the periphery of the geometry to represent the modeling of a heat pipe in the reactor.
For simplicity, this region only represents the heat pipe vapor core, no heat pipe wall is represented in the geometry.
The modeling approach used in this demonstration problem follows the one used in \citet{price2023thermal} and \citet{price2024multiphysics}.
Neutronically, the heat pipe region acts as a void as the vapor core of a heat pipe has a negligible neutronic cross section.
For the conduction problem, the heat pipe can be modeled using a Robin boundary condition with no explicit modeling of heat conduction within the vapor core region.
This boundary condition is imposed on the interface between Material B and the voided region.
To create a nontrivial $\breve{T}_i(\bm{r})$, the boundary condition imposed on the edge of this vapor core region is specified with $A_2 = 1\,\frac{\text{cm}}{\text{K}}$, $B_2 = 8 \, \frac{1}{\text{K}}$ and $C_2(\bm{r}) = 3000 \cos(6\varphi)$.
In this expression, $\bm{r}= (r,\varphi)$ is given in a polar coordinate system centered on the 30$^\circ$ vertex of the triangular pin cell.
With this boundary condition, MOSS is used only to calculate the source-driven component $\tilde{T}_i(\bm{r})$ within the solid regions while the boundary-driven harmonic component $\breve{T}_i(\bm{r})$, determined by the inhomogeneous Robin boundary condition, is supplied by the MOOSE heat transfer module.
A reflective boundary condition is imposed for both processes along the outside of the pin cell to create the repeating hexagonal geometry.
This particular problem is selected to demonstrate that, even though MOSS provides a unified approach to solving the heat conduction and neutron transport equations, differing physical processes often require different geometries leading to further complications in history tracking in the current form of MOSS.

For this demonstration problem, a more complex branching process is required for particles crossing the voided boundary.
Each particle must branch separately into a transmitted neutron and, with $\lvert \alpha^{(h)}_2\rvert = 0.67$ survival probability, a reflected heat particle.
This introduces an additional branching process in addition to the absorption branching shown in the previous problem.
This now also leads to an additional category of dedicated computational effort.
Namely, previous computational effort could be divided into effort spent solely on the heat conduction problem and effort spent simultaneously solving the neutron transport and heat conduction.
Now, computational effort can be spent solely on the neutron transport problem as well.

\begin{figure}[htp]
    \centering
    \includegraphics[width=\textwidth]{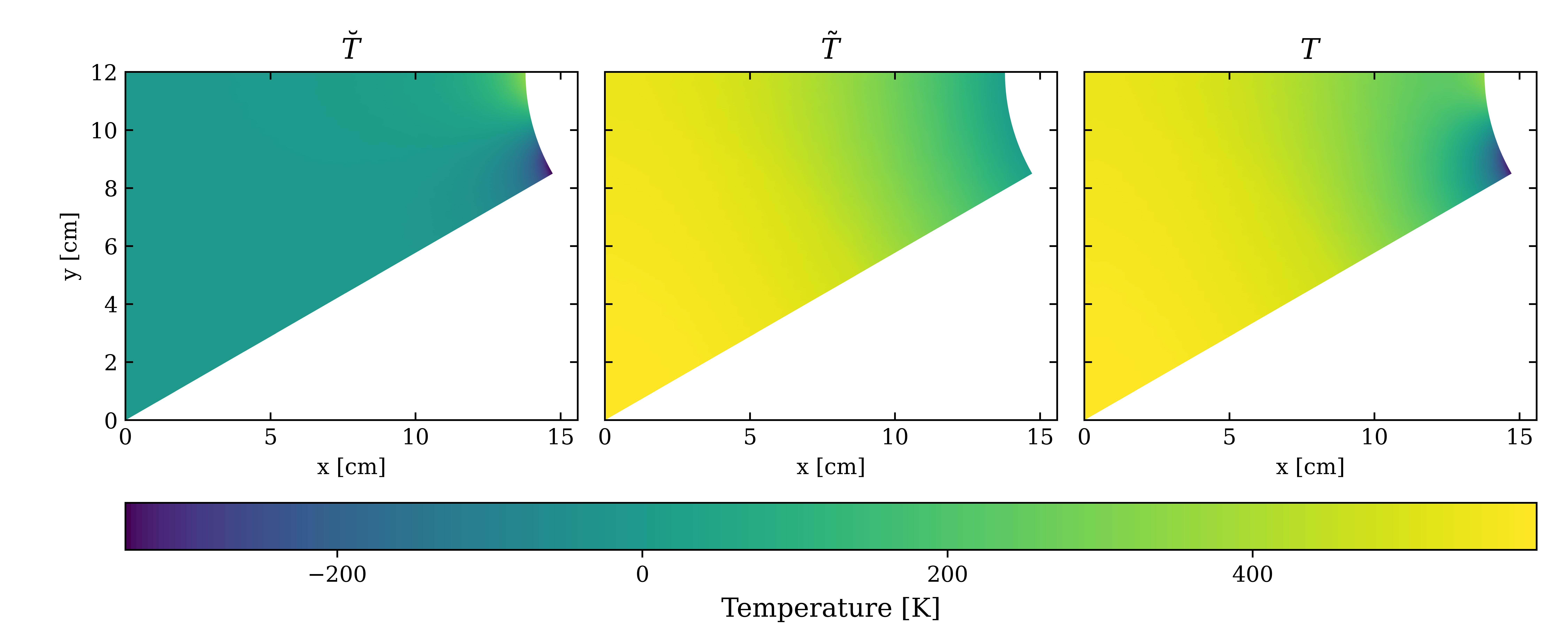}
    \caption{Reference solutions for the temperature distributions associated with the pin cell demonstration problem. Calculated using the finite element method through the MOOSE heat transfer module. In this study, MOSS is used to calculate $\tilde{T}$ which can be combined with $\breve{T}$ to form the full solution $T$. \hl{The results shown in this figure are used to verify MOSS.}}
    \label{fig:temperature_reference}
\end{figure}

\begin{figure}[htp]
    \centering
    \includegraphics[width=.65\textwidth]{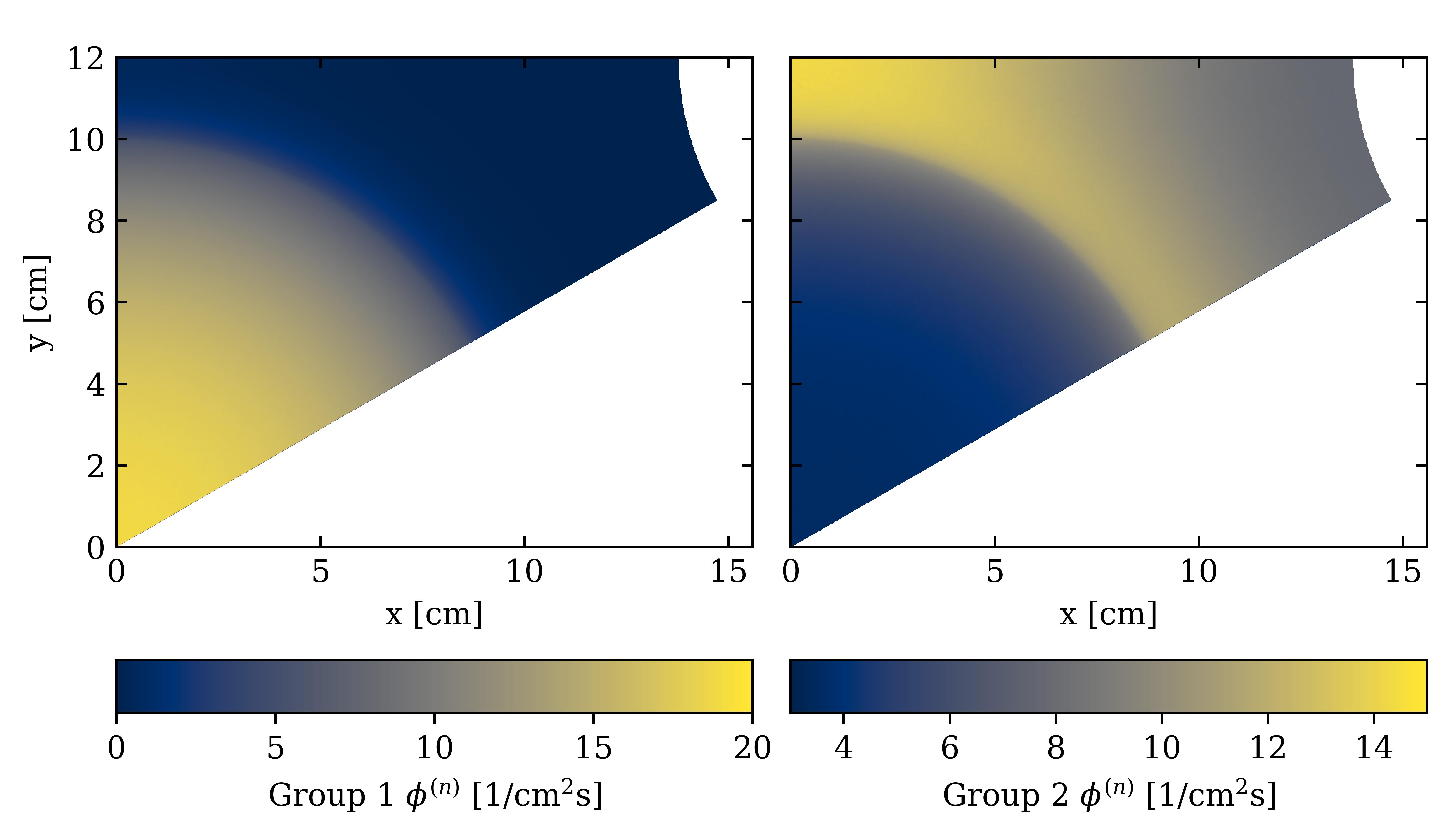}
    \caption{Reference solutions for the 2-group flux distributions associated with the pin cell demonstration problem. Calculated using OpenMC \hl{to verify the flux distributions produced by MOSS}.}
    \label{fig:flux_reference}
\end{figure}

The reference values for the temperature and flux solutions are shown in Figure \ref{fig:temperature_reference} and Figure \ref{fig:flux_reference}, respectively.
The reference temperature distributions are calculated using the finite element method in the MOOSE framework and the reference flux distributions are calculated using OpenMC.
Being a demonstration problem, negative temperatures are acceptable.
In Figure \ref{fig:temperature_reference}, all variability in the temperature distribution associated with the boundary-driven harmonic component, $\breve{T}$, arises from the imposition of inhomogeneous boundary conditions.
MOSS is used only to compute the source-driven component $\tilde{T}$ that arises from the presence of the source.
\hl{Despite the possibility for discretization-induced errors arising from the application of the finite element method, the values shown in Figure} \ref{fig:temperature_reference} \hl{are considered exact as these errors end up being significantly smaller in magnitude than those introduced by the transport approximation to the heat conduction equation.}

\begin{figure}[htp]
    \centering
    \includegraphics[width=.9\textwidth]{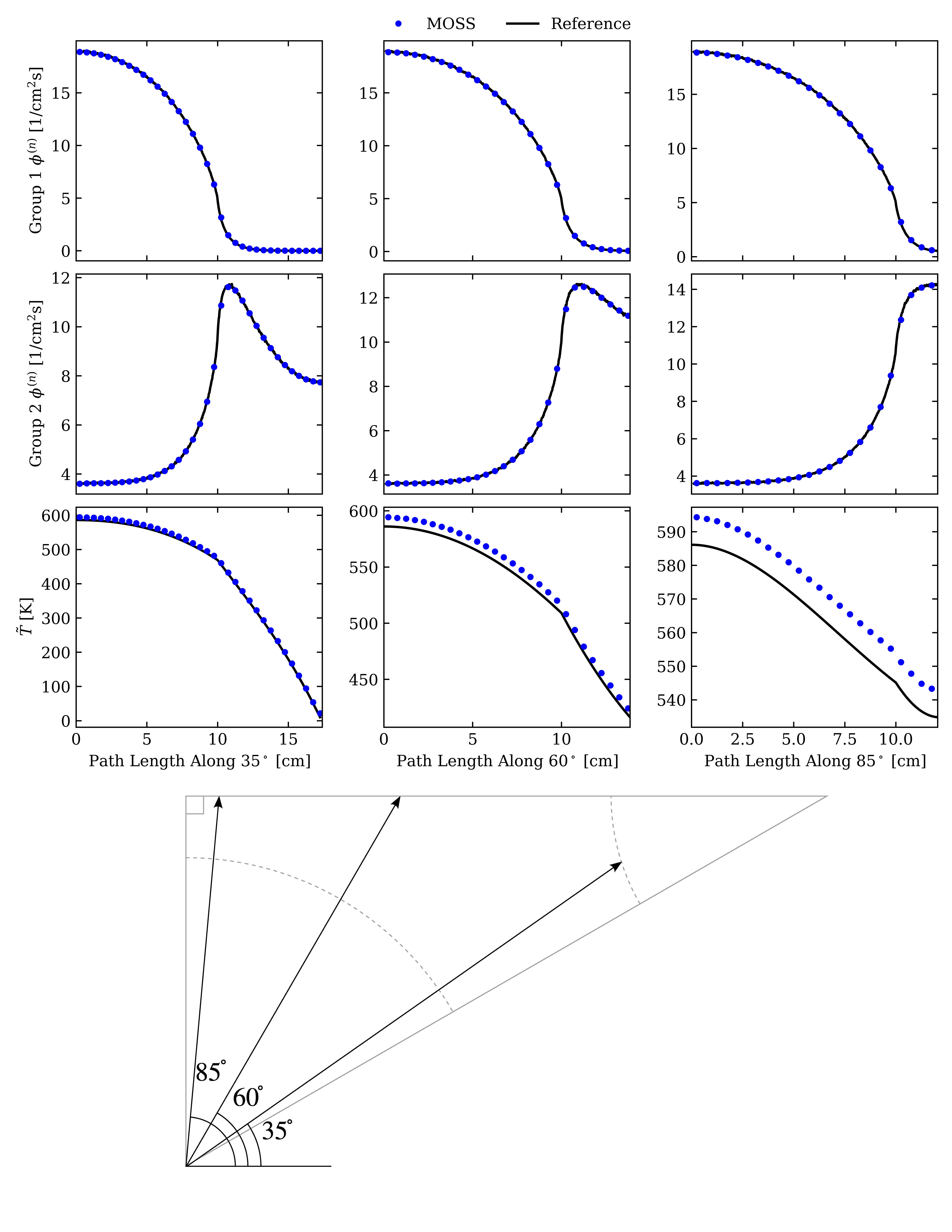}
    \caption{Reference versus MOSS computed values for temperature and flux distribution along multiple rays traversing the pin cell geometry (top). Orientation of rays intersecting origin described by angle from x$^+$ axis (bottom).}
    \label{fig:full_page}
\end{figure}

Figure \ref{fig:full_page} compares solutions calculated with MOSS to the reference values for multiple ray traversals of the pin cell geometry.
Again, a large number of samples are used to ensure the stochastic errors arising strictly from the Monte Carlo method are negligible in this figure.
The OpenMC-calculated reference neutron flux agrees with the neutron flux calculated by MOSS.
\hl{For the temperature distribution, there is disagreement between the MOSS and reference temperature distribution due to the transport approximation.}
Both of these observations are consistent with those made in the slab demonstration problem.

\begin{figure}[htp]
    \centering
    \includegraphics[width=.65\textwidth]{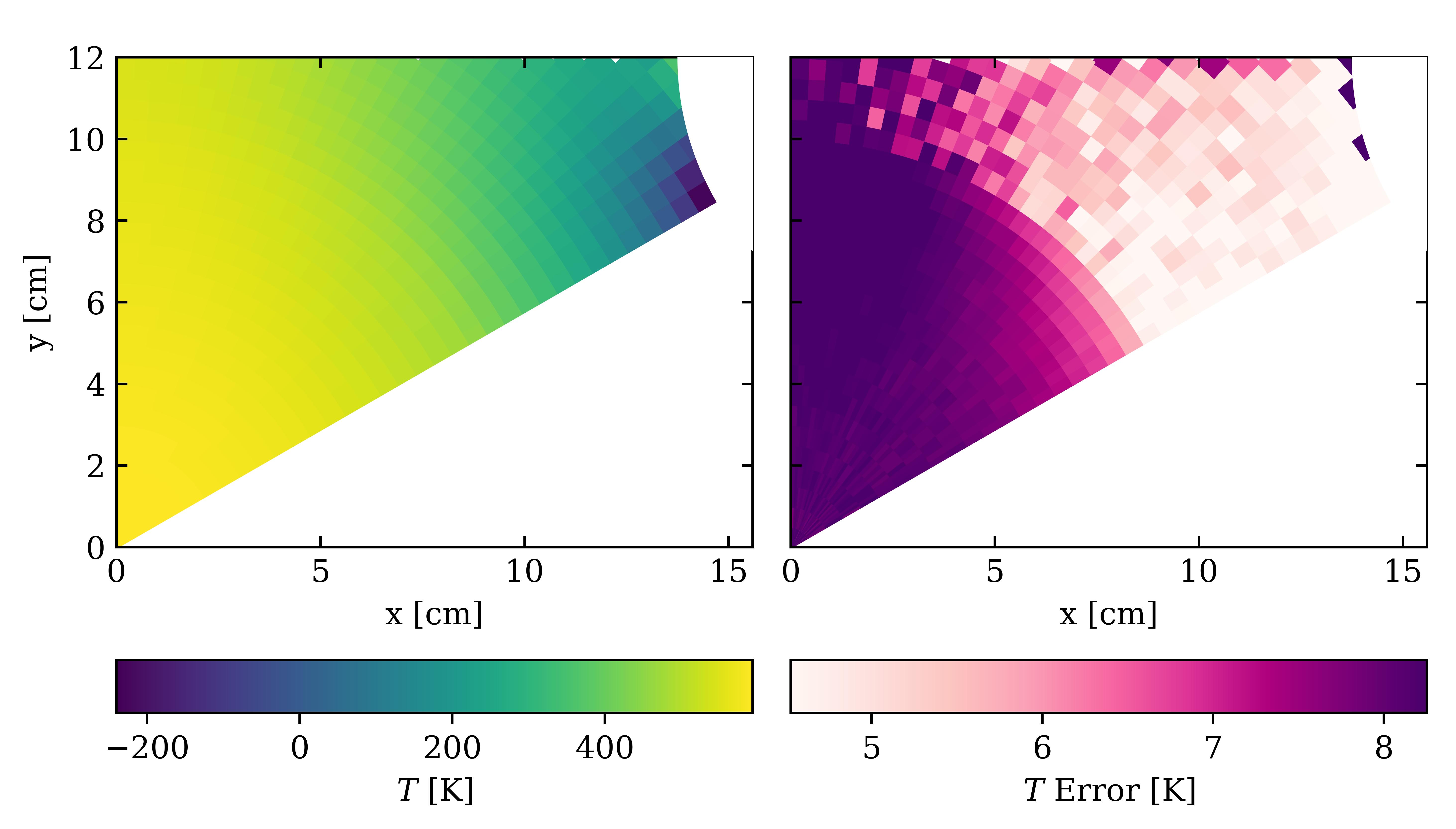}
\caption{\hl{Full temperature distribution for the pin cell demonstration problem, with} \(\tilde{T}\) \hl{calculated by MOSS and }\(\breve{T}\) \hl{supplied by MOOSE (left). The absolute error relative to the full MOOSE solution is shown on the right. The error includes both Monte Carlo statistical uncertainty and the bias introduced by the transport approximation to the heat conduction equation.}}    \label{fig:tfinal}
\end{figure}

\hl{Finally, the calculated source-driven component} $\tilde{T}$ \hl{can be combined with the MOOSE-calculated boundary-driven harmonic component} $\breve{T}$ \hl{to form the full temperature field,} $T$.
Figure \ref{fig:tfinal} shows the disagreement between $T$ calculated partially with MOSS and $T$ calculated entirely by the finite element method.
Although the statistical uncertainty arising from the Monte Carlo calculation method varies largely over the problem geometry due to differing tally region sizes and sampling densities, the trends are still clear with the MOSS-calculated temperature distributions overestimate the reference.
\hl{This results in a mean absolute error of about 7.4 K}.
Aside from this stochastic error source, the errors observed at this scale can be attributed entirely to approximating heat conduction in the calculation of $\tilde{T}$ as shown in Figure \ref{fig:full_page}, as opposed to the calculation of $\breve{T}$.
With this, the trends observed are consistent with those shown in Figure \ref{fig:full_page} where larger errors are observed local to the source region.

\section{Conclusion}
\label{sec:conc}

MOSS was introduced as a Monte Carlo based method that uses analytically-derived weighting factors to track the solution to the heat conduction equation while solving the neutron transport equation.
\hl{In this presentation, MOSS can only be used directly to solve for the source-driven component with homogeneous boundary conditions.
Therefore, the finite element method was used to supply the boundary-driven harmonic component of the temperature distribution associated with inhomogeneous boundary conditions.}
Two demonstration problems were presented that relied on the same set of materials: (1) a two-region slab and (2) a two-region hexagonal pin cell problem.
Some main drawbacks of MOSS were observed in these demonstration problems.
Specifically, branching led to computational time dedicated solely to solving one physics, and biases were introduced from the approximation of heat conduction as a Boltzmann transport process.
The \hl{material properties} had to be carefully selected so that the theoretical heat particle stochastic process remained sufficiently similar to the neutronic process, reducing the risk of large or unbounded variance in the estimators used to calculate the temperature distribution.

There are a large number of areas for expansions beyond the current work.
For example, the random walk and branching process can be more carefully tuned to maximize the amount of computational time dedicated to simultaneously solving both physics.
Additionally, the inclusion of an adjustable particle weight corresponding to the neutronics process could enable more flexibility for further improvements.
There is also some room for future work in the theoretical development of the method.
Some generalizations can be presented that allow for the treatment of non-identical source distributions, alternative boundary conditions and eigenvalue problems.
The potential for cross-physics feedback should also be explored.
For example, in coupled problems, estimations of the temperature distribution could be fed into the neutronic random walks by adjusting the cross sections accordingly.
Secondary heating could also be accounted for by increasing heat particle weights on neutronic interactions.
Finally, the use of MOSS for eigenvalue problems with multiphysics feedback could yield efficient workflows as \hl{the source-driven component of the temperature distribution} could be evolved alongside the fission source distribution.
Regardless, this study represents a first step towards further explorations of MOSS and its applications.

%\section*{Acknowledgment}

\appendix
\input{appendix}

\bibliographystyle{elsarticle-num-names} 
\bibliography{refs.bib}

\end{document}

%% file: appendix.tex
\section{Violations of the Central Limit Theorem}
\label{ap:clt_vio}

\hl{If the stochastic processes governed by the heat transfer and neutron transport physics become too dissimilar, the variance associated with the statistical estimators for the temperature distribution can become unbounded.}
\hl{As discussed in Section} \ref{sec:clt_vio}\hl{, this creates a situation where the usual central limit theorem assumptions are not satisfied and the nominal Monte Carlo uncertainty estimates may no longer describe convergence.}
\hl{In MOSS, this issue is tied to the correlated sampling weights used to reinterpret neutron histories as heat-particle histories.}
\hl{When the two processes are closely matched, these weights remain well behaved and the estimator maintains the expected behavior.}
\hl{When they are poorly matched, rare histories can carry disproportionately large heat-particle weights, causing the apparent RMSE to be dominated by sampling of those rare events rather than by the usual gradual reduction in statistical uncertainty.}

\begin{figure}[htp]
    \centering
    \includegraphics[width=.5\textwidth]{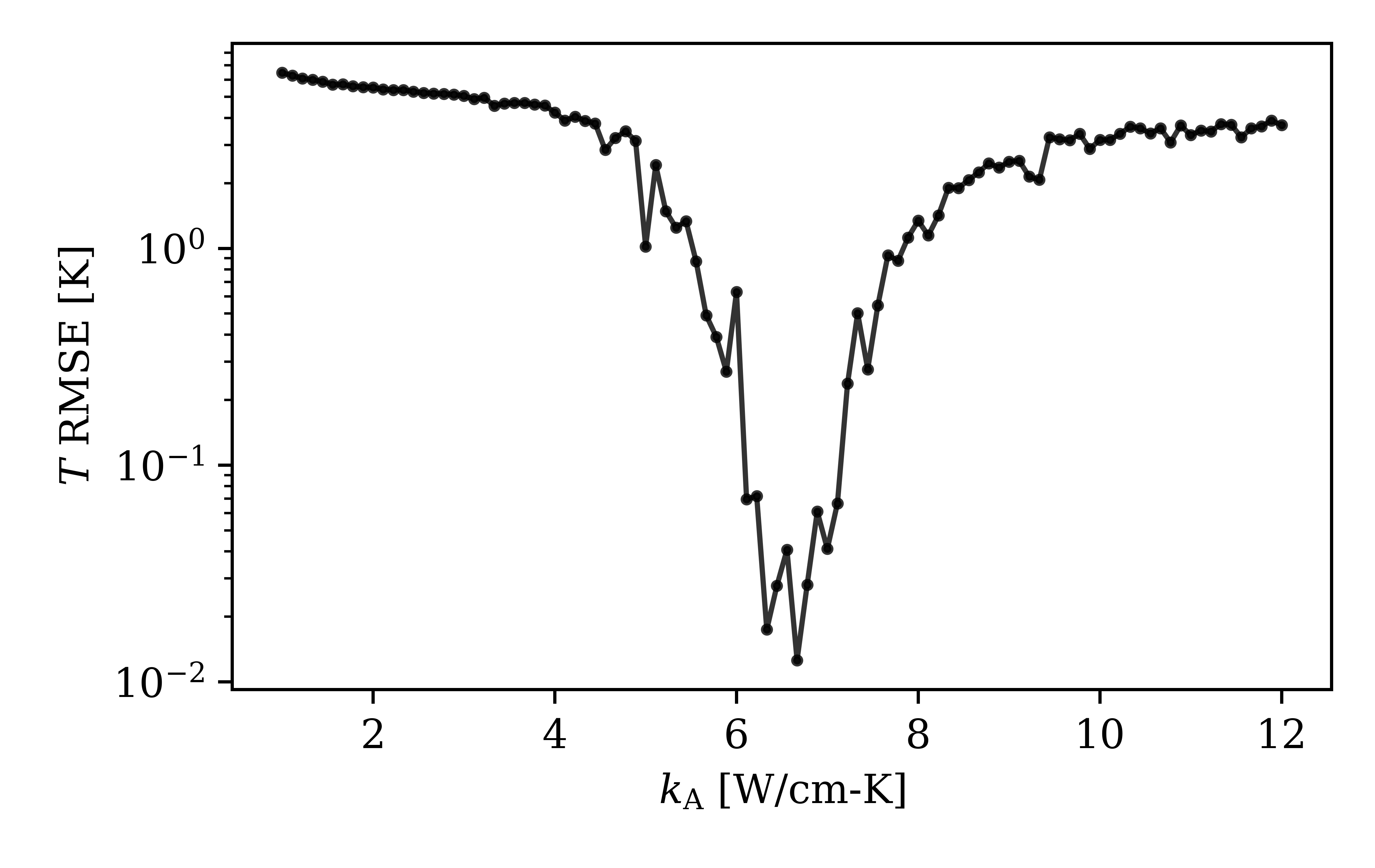}
    \caption{\protect\hl{Temperature RMSE from MOSS for the slab demonstration problem as Material A thermal conductivity,} $k_A$\protect\hl{, varies. The increase in error away from the minimum illustrates variance growth caused by mismatch between the neutron and heat-particle stochastic processes.}}
    \label{fig:clt_divergence}
\end{figure}

\hl{To illustrate the prominence of this effect, Figure} \ref{fig:clt_divergence} \hl{shows the temperature RMSE for the slab demonstration problem over a sweep of the Material A thermal conductivity,} $k_A$\hl{.}
\hl{All other material properties and problem settings are left unchanged from the slab case in Section} \ref{sec:resultss}\hl{.}
\hl{The minimum occurs near the value used in Table} \ref{tab:xsecs}\hl{, where the neutronic and heat-particle stochastic processes were intentionally selected to be similar.}
\hl{Away from this region, the RMSE increases significantly indicating that the variance of the estimator is growing.}
\hl{This demonstrates why the material properties in the demonstration problems were selected carefully and why future applications of MOSS require variance-control strategies when the two stochastic processes are not naturally aligned.}

\section{Convergence Verification}
\label{ap:cv}
\begin{figure}[htp]
    \centering
    \includegraphics[width=\textwidth]{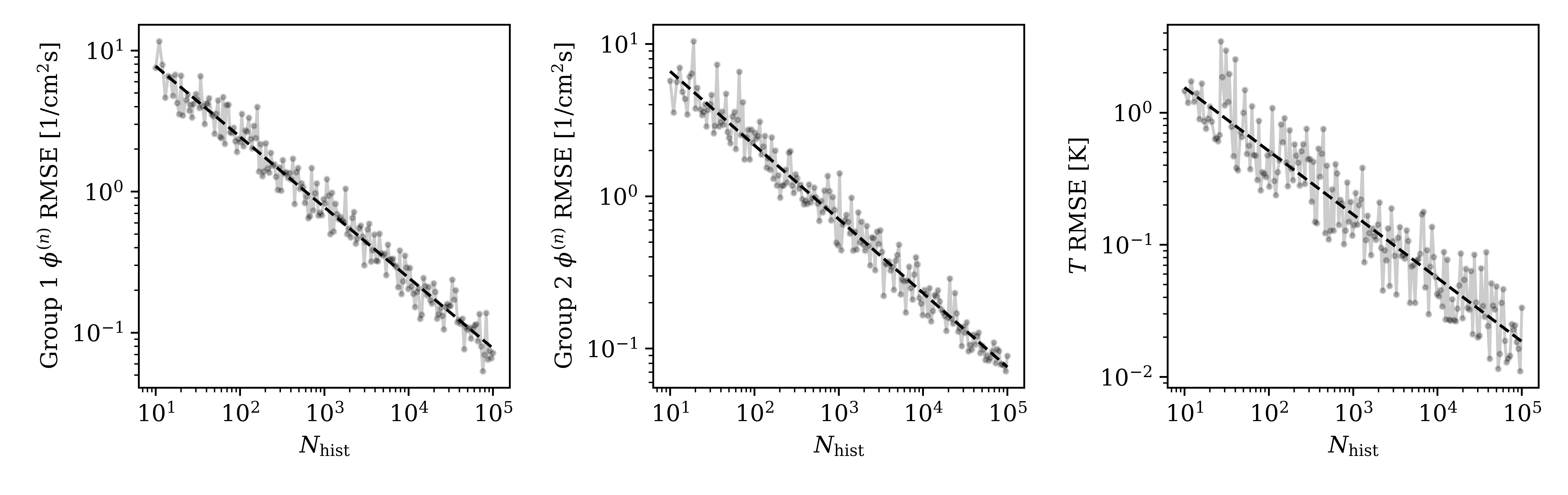}
    \caption{\protect\hl{Results from the convergence study performed on the slab demonstration problem to verify MOSS. Gray points indicate RMSE values evaluated using different numbers of histories,} $N_{\text{hist}}$\protect\hl{, for the group-1 neutron scalar flux,} $\phi^{(n)}_{1}$\protect\hl{, the group-2 neutron scalar flux,} $\phi^{(n)}_{2}$\protect\hl{, and the temperature,} $T$\protect\hl{. Dashed lines show log-log regression fits with slopes of} $-0.50$\protect\hl{,} $-0.49$\protect\hl{, and} $-0.48$\protect\hl{, respectively, consistent with the expected Monte Carlo convergence rate.}}
    \label{fig:convv}
\end{figure}

\hl{A fundamental limitation of treating the heat transport problem with a Monte Carlo method is that the characteristic} $\sqrt{N_{\text{hist}}}$ \hl{convergence rate is inherited by a problem that is ordinarily solved using deterministic methods with substantially more favorable convergence properties.}
\hl{Here,} $N_{\text{hist}}$ \hl{is the total number of histories run.}
\hl{To illustrate this limitation and verify the method discussed here, Figure} \ref{fig:convv} \hl{shows the results of a convergence study performed on the slab demonstration problem.}
\hl{The results from Figure} \ref{fig:ref} \hl{are used for comparison when calculating root mean squared error (RMSE) and are the result of using around} $8 \times 10^7$ \hl{histories.}
\hl{The} $\sqrt{N_{\text{hist}}}$ \hl{convergence rate is reflected by the roughly} $-\frac{1}{2}$ \hl{slope of the regression line in log-transformed space.}

\section{Optical Thickness and the Transport Approximation}
\label{ap:optthic}
This section provides a brief exploration of how the optical thickness of a problem impacts the accuracy of the transport approximation applied to the heat conduction equation.
It will demonstrate that optically thick, or low thermal conductivity, problems yield a more accurate transport approximation.
In this section, MOSS is not used.
Instead, consistent with the single-physics approaches demonstrated in \citet{Fraley1977MonteCarlo} and \citet{cho2010monte}, the conduction process is treated alone with the random walks directly following the analogous transport process.
As estimators derived from the correlated sampling approach underlying MOSS are unbiased, any disagreement observed in this single-physics demonstration is directly applicable to the expected disagreements between MOSS and the non-transport temperature solution.
Although $\beta=1$ is maintained for this section, it can be modified to artificially increase the optical thickness of a problem.

\begin{figure}[htp]
    \centering
    \includegraphics[width=.5\textwidth]{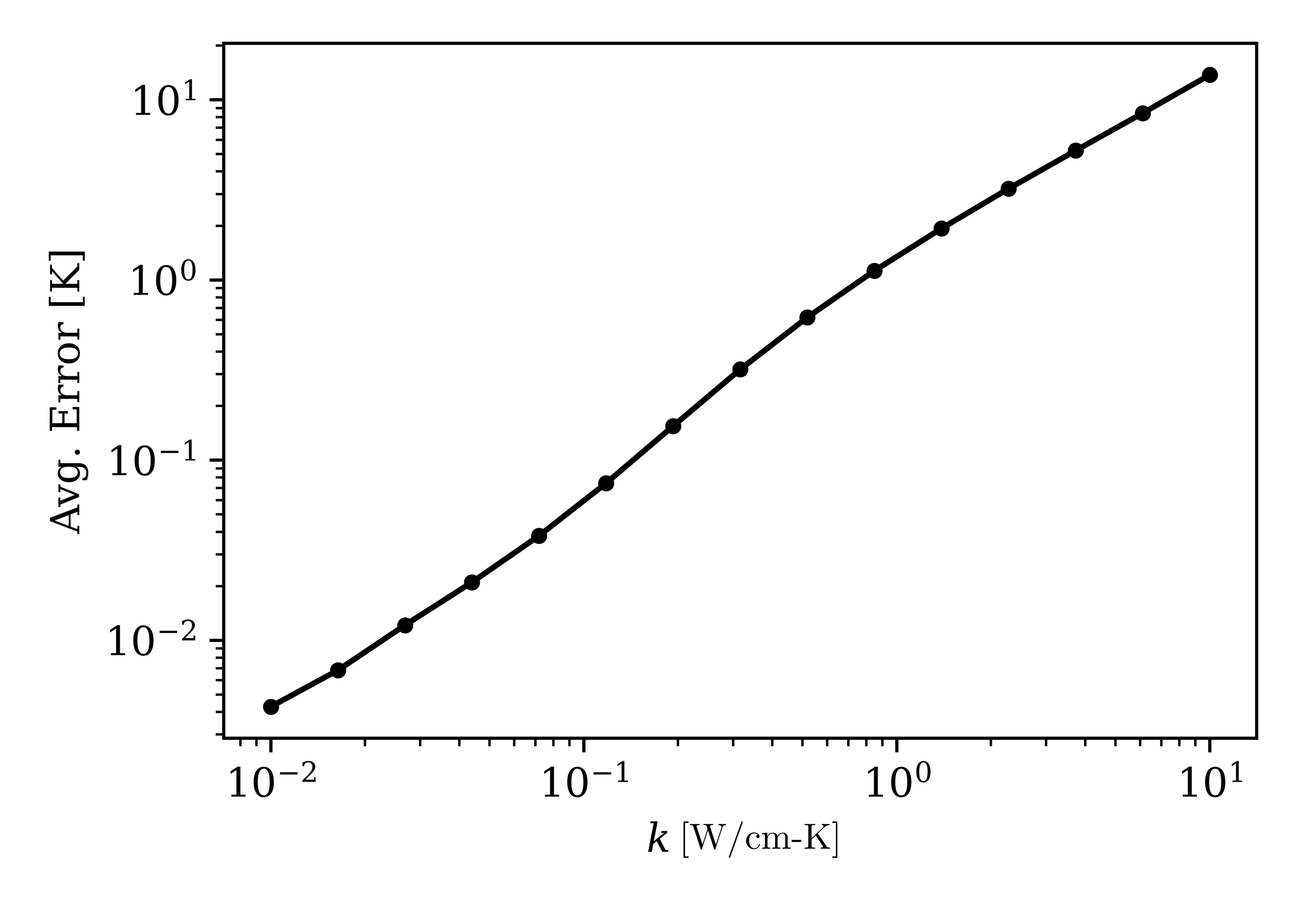}
    \caption{Average error introduced strictly from the transport approximation of the conduction equation as a function of thermal conductivity.}
    \label{fig:kerrs}
\end{figure}

For simplicity, a one-region one-dimensional slab geometry will be treated with a uniform heat source.
The homogeneous boundary condition $A_i = 1\, \frac{\text{cm}}{\text{K}}$, $B_i = 2\, \frac{1}{\text{K}}$ and $C_i(\bm{r})=0$ will be applied to both boundaries.
For some $Q^{(h)}$ as the volumetric heat generation rate, thermal conductivity $k$ and origin-centered region with width $1\, \text{cm}$, the analytically derived temperature distribution for this single region is
\begin{gather}
T(x) = \frac{1}{8}\frac{Q^{(h)}}{k}\left(3-4x^2 \right).
\end{gather}
In Figure \ref{fig:kerrs}, spatially averaged errors from approximating this solution for a variety of values of $k$ using the Monte Carlo method applied to the transport approximation are shown.
For each different $k$ shown in this figure, $Q^{(h)}$ is modified proportionally to maintain $\frac{Q^{(h)}}{k}=1 \frac{\text{K}}{\text{cm}^2}$ such that the magnitude of the solution does not change.
In this figure, increasing $k$ decreases the optical thickness associated with the analogous transport process.
As expected, this makes the conduction process less ``transport-like'' and reduces the accuracy of the transport approximation.

